\documentclass[prb,twocolumn,superscriptaddress,showpacs]{revtex4}

\usepackage{graphicx}
\usepackage{amssymb}
\usepackage{amsmath}

\begin{document}

\title{In-plane radiative recombination channel of a dark exciton in self-assembled quantum~dots}

\author{T.~Smole\'nski}
 \email{Tomasz.Smolenski@fuw.edu.pl}
\author{T.~Kazimierczuk}
\author{M.~Goryca}
\author{T.~Jakubczyk}
\affiliation{%
Institute of Experimental Physics, Faculty of Physics, University of
Warsaw,\\
ul. Ho\.za 69, 00-681 Warsaw, Poland }%
\author{\L{}.~K\l{}opotowski}
\author{\L{}.~Cywi\'nski}
\author{P.~Wojnar}
\affiliation{%
Institute of Physics, Polish Academy of Sciences,\\
Al. Lotnik\'ow 32/64, 02-688 Warsaw, Poland}%
\author{A.~Golnik}
\author{P.~Kossacki}
\affiliation{%
Institute of Experimental Physics, Faculty of Physics, University of
Warsaw,\\
ul. Ho\.za 69, 00-681 Warsaw, Poland }%

\begin{abstract}

We demonstrate evidence for a radiative recombination channel of dark excitons in self-assembled quantum dots. This channel is due to a light hole admixture in the excitonic ground state. Its presence was experimentally confirmed by a direct observation of the dark exciton photoluminescence from a cleaved edge of the sample. The polarization resolved measurements revealed that a photon created from the dark exciton recombination is emitted only in the direction perpendicular to the growth axis. Strong correlation between the dark exciton lifetime and the in-plane hole g-factor enabled us to show that the radiative recombination is a dominant decay channel of the dark excitons in CdTe/ZnTe quantum dots.

\end{abstract}


\maketitle

Self-assembled semiconductor quantum dots (QDs) are recognized as a medium for storage and manipulation of quantum information \cite{loss_tsmol_prl, liu_tsmol_prl}. Important applications, involving emission of single photons \cite{single_photon_tsmol_prl} or entangled photon pairs \cite{entangled_photon1_tsmol_prl, entangled_photon2_tsmol_prl} rely heavily on properties of confined excitonic complexes. In particular, the central point of many schemes is a neutral exciton consisting of a single electron and a single hole. However, the properties of such a complex depend on the relative orientation of the angular momenta of confined carriers. An electron and a hole with antiparallel angular momenta form an optically active \emph{bright exciton} ($\mathrm{X}_\mathrm{b}$) with total angular momentum projection on the QD growth axis $J_z=\pm 1$, while parallel orientation of angular momenta leads to the formation of a \emph{dark exciton} ($\mathrm{X}_\mathrm{d}$) with $J_z=\pm 2$. Due to the lack of dipole allowed recombination channel, the lifetime of the latter may extend above microseconds \cite{labeau_tsmol_prl}. The presence of the dark excitons is often neglected \cite{suffczynski_tsmol_prl}, however due to their persistent nature, dark excitons can have detrimental effect on the properties of QD devices. On the other hand, recent findings show that the dark exciton can be also used as a qubit \cite{gershoni_tsmol_prl}, which turns its long lifetime into an advantage. The important role of the dark excitons and their lifetime is therefore of principal interest for quantum information processing.

Studies of dark excitons are hindered by their absence in the photoluminescence (PL) spectrum. Dark excitons become visible after introducing a mixing between bright and dark configurations, either by an in-plane magnetic field \cite{dark_cdse_tsmol_prl, bayer_tsmol_prl} or by the exchange interaction with magnetic dopant \cite{goryca_tsmol_prl}. The lifetime of the dark exciton is widely believed to be determined by a spin-flip process turning a dark exciton into a bright one \cite{labeau_tsmol_prl, dalgarno_tsmol_prl, smith_tsmol_prl, roszak_tsmol_prl}. The main argument for such a mechanism is a biexponential decay of the bright exciton, which was observed in both III-V and II-VI QDs \cite{johansen_tsmol_prl, labeau_tsmol_prl}. However, the spin-flip cannot be considered the sole decay mechanism of the dark exciton. It is especially evident in the analysis of the low temperature behavior of this process. In particular, turning the dark exciton into a bright one requires absorption of energy from the phonon bath equal to (isotropic) electron-hole exchange constant $\delta_0$ \cite{bayer_tsmol_prl}, and therefore predicted lifetime of the dark exciton can, in principle, be infinitely prolonged by lowering the temperature. A presence of another decay mechanism was already suggested in Ref. \onlinecite{johansen_tsmol_prl}, where the authors concluded that the lifetime of the dark exciton is limited by an unknown ``non-radiative decay channel'', rather than by the spin-flip process.

In this Communication, we demonstrate that the dark exciton can recombine radiatively, emitting linearly polarized light in a direction perpendicular to the growth axis. This fact, together with our observation of a strong quantitative correlation between the $\mathrm{X_d}$ radiative lifetime and the in-plane hole g-factor, proves that this recombination channel is due to an admixture of the light hole component in the hole ground state. We also provide an evidence that the radiative recombination can be dominant decay channel of the dark exciton in CdTe/ZnTe QDs. However, our findings should apply also to other QDs exhibiting heavy-light hole mixing.

Our experiments were performed on samples containing self-assembled CdTe/ZnTe QDs grown by the amorphous tellurium desorption method \cite{tinjod_tsmol_prl}. The measurements were carried out in a micro-photoluminescence setup described in Refs. \onlinecite{kazimierczuk_2010_tsmol_prl} and \onlinecite{kazimierczuk_2011_tsmol_prl}. Sample was placed inside either a continuous-flow or a magneto-optical bath cryostat at temperature 1.5-10K. During the time-resolved measurements the QDs were excited nonresonantly by a frequency doubled pulses from a Ti:sapphire laser. The laser repetition rate was effectively reduced with a pulse picker, which enabled us to observe long PL decays. The PL was recorded either by a CCD camera or, in case of time-resolved measurements, by an avalanche photodiode (APD) with sub-nanosecond temporal resolution.

\begin{figure}
\includegraphics{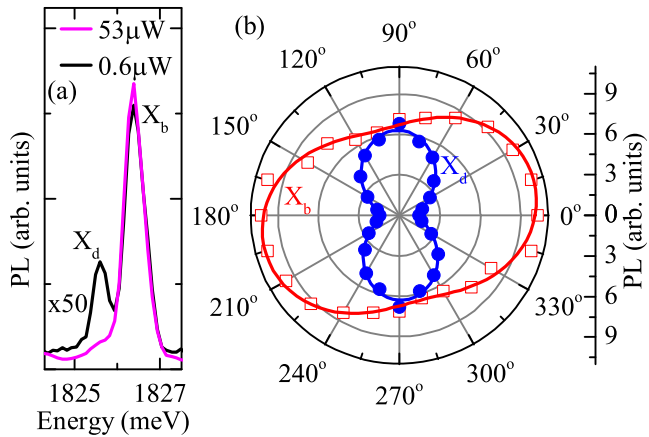}
\caption{(color online)
PL of the neutral exciton measured from the cleaved edge of the sample at zero magnetic field under CW excitation.
(a) PL spectra measured under different excitation powers related to different average dot repopulation times (the saturation power of this QD corresponds to 2mW). Relative intensity of the $\mathrm{X_d}$ line becomes stronger for the repopulation time comparable to dark exciton lifetime. (b) Polar plot presenting the polarization of the dark and bright exciton emission for $0.6\mu$W excitation power. The $90^\circ$ and $270^\circ$ directions are along the QD growth axis. Symbols for angles $\varphi$ and $\varphi+180^\circ$ correspond to the same data points.
 \label{fig1}}
\end{figure}

In this study we investigate the presence of a radiative emission of the dark exciton without external magnetic field. Such an emission is not observed in PL measurements performed in a typical experimental setup with optical axis parallel to the QD growth axis \cite{bayer_tsmol_prl,kazimierczuk_2011_tsmol_prl,leger_tsmol_prl}. However, the dark exciton PL can be revealed by using an in-plane setup geometry (i.e., the optical axis perpendicular to the growth axis). Figure \ref{fig1}(a) presents PL spectra of the neutral exciton measured in such a geometry under nonresonant, continuous wave excitation. Under strong excitation power (short dot repopulation time) the bright exciton PL dominates the spectrum, as $\mathrm{X_b}$ lifetime is usually much shorter compared to the dark exciton one. However, in the low excitation power regime (i.e., when the dot repopulation time is long enough to allow undisturbed recombination of the $\mathrm{X_d}$), the dark exciton emission line is clearly visible (its identification is based on the spectral position, which matches relative energy of the dark exciton determined in the magneto-photoluminescence experiments\cite{kazimierczuk_2011_tsmol_prl}). This result indicates the presence of \emph{radiative recombination channel} of the dark exciton confined inside a QD. Until now, such an effect was demonstrated only in [111]-grown pyramidal QDs \cite{karlson_tsmol_prl,dupertuis}. We emphasize that dark exciton PL could be discovered only using the in-plane geometry. It is related to the fact that $\mathrm{X_d}$ emission line is almost fully linearly polarized along the QD growth axis (Fig. \ref{fig1}(b)).

In order to understand the origin of the observed radiative $\mathrm{X_d}$ decay mechanism, one has to take into account the valence band mixing, which is usually present in both III-V \cite{krizankovski_tsmol_prl,atature_tsmol_prl,ohno_tsmol_prl, karlson_tsmol_prl} and II-VI \cite{goryca_tsmol_prl, leger_tsmol_prl, koudinov_tsmol_prl, cao_tsmol_prl} QDs. We consider the heavy ($|\pm3/2\rangle$) and the light ($|\pm1/2\rangle$) hole states, which are mixed due to the presence of a QD shape or strain anisotropy. In the leading order, $|\pm3/2\rangle$ states are only mixed with $|\mp1/2\rangle$ \cite{leger_tsmol_prl}, which yields the lowest-energy hole states in a QD to be given by $|\phi_{h\pm}\rangle=|\pm3/2\rangle+\epsilon_{\pm}|\mp1/2\rangle$ ($\epsilon_{\pm}=\epsilon e^{\pm 2 i \theta}$, where $\epsilon$ represents the strength of the valence band mixing, and $\theta$ its direction). The ground states of the dark exciton are now of the forms $|\uparrow_e\rangle|\phi_{h+}\rangle$, $|\downarrow_e\rangle|\phi_{h-}\rangle$, where $|\uparrow_e\rangle$, $|\downarrow_e\rangle$ denote the electron spin eigenstates. One can define the hole states using the $X,Y,Z$ $p$-type orbitals and $|\uparrow_h\rangle$, $|\downarrow_h\rangle$ hole spin eigenstates \cite{leger_tsmol_prl}. Thus, each of the dark exciton states can be expressed in the following way
\begin{multline}
\label{valence_band}
|\uparrow_e\rangle|\phi_{h+}\rangle=-\frac{1}{\sqrt{2}}|\uparrow_e\rangle|\uparrow_h\rangle|X+iY\rangle\\
+\frac{1}{\sqrt{6}}\epsilon_{+}|\uparrow_e\rangle|\uparrow_h\rangle|X-iY\rangle+\sqrt{\frac{2}{3}}\epsilon_{+}|\uparrow_e\rangle|\downarrow_h\rangle|Z\rangle,
\end{multline}
($|\downarrow_e\rangle|\phi_{h-}\rangle$ is given by an analogous expression).
Within our convention of the hole spin, the radiative recombination requires antiparallel orientation of the electron and the hole spin. Therefore, the admixtures represented by the first two terms in Eq. (\ref{valence_band}) do not allow radiative recombination, as the electron and the hole spins are parallel. However, the admixture represented by the last term $\epsilon_{+}\sqrt{2/3}|\uparrow_e\rangle|\downarrow_h\rangle|Z\rangle$ in Eq. (\ref{valence_band}) has antiparallel spin directions of the carriers, which opens the radiative recombination channel of the dark exciton. Due to the presence of the $Z$ orbital in the mentioned term, the dipole moment of the $\mathrm{X_d}$ optical transition is oriented along the QD growth axis ($z$). Such a result fully confirms our experimental observation that the dark exciton PL is directed perpendicular to the growth axis\cite{karlson_tsmol_prl}. Within our model, the oscillator strengths of the dark exciton optical transitions in each of $|\uparrow_e\rangle|\phi_{h+}\rangle$ and $|\downarrow_e\rangle|\phi_{h-}\rangle$ states are equal and proportional to $\epsilon^2$. However, one should remember that due to the QD anisotropy, the dark exciton eigenstates are usually linear combinations of the above states, given by $|\uparrow_e\rangle|\phi_{h+}\rangle \pm e^{i\varphi} |\downarrow_e\rangle|\phi_{h-}\rangle$. The behavior of these states exhibits complex dependence on the relative orientations of the QD anisotropy ($\delta_2$ splitting of the dark excitons) and the valence band mixing. E.g., in the pure $C_{2v}$ QD shape symmetry, only one of the dark excitons has a non-vanishing dipole transition moment \cite{dupertuis}. Detailed discussion of these effects remains beyond the scope of this Communication. We note only that any given combination of relative orientation of involved interactions leads to the $\epsilon^2$ dependence of the dark exciton radiative recombination rate.
\begin{figure}
\includegraphics{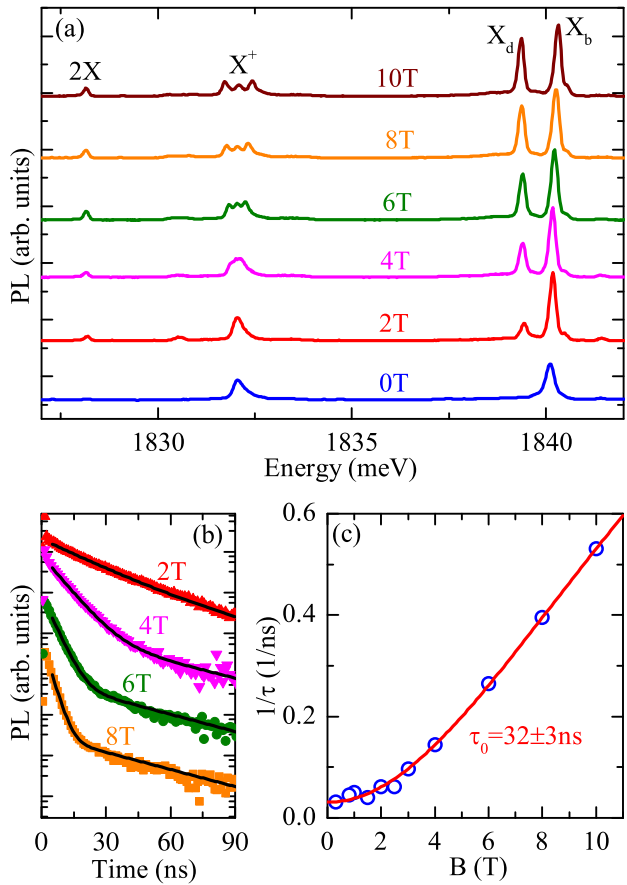}
\caption{(color online) (a) PL spectra of a single QD for different values of the in-plane magnetic field. (b) PL decay curves of the dark exciton for different magnetic fields, measured without polarization resolution. Solid lines represent the fits of biexponential decays, which are related to the presence of two dark excitons with almost equal emission energies, but different efficiencies of mixing with bright excitons. (c) Inverse lifetime of the shorter lived dark exciton as a function of the in-plane magnetic field. The solid line represents the fitted curve described by Eq. (\ref{dark_oscilator_stength}). The zero magnetic field lifetime of the dark exciton for this QD yields $\tau_0=32\pm 3$ns.
 \label{fig2}}
\end{figure}

In the following, we show that in the case of CdTe/ZnTe QDs, the $\mathrm{X_d}$ recombination is indeed limited by a process with a rate proportional to $\epsilon^2$. In order to do so, we experimentally extract the $\mathrm{X_d}$ recombination rate in a time-resolved PL measurement. On the other hand, a convenient measure of the valence band mixing strength $\epsilon$ is a value of the in-plane hole g-factor $g_h$, directly proportional to the mixing strength\cite{koudinov_tsmol_prl}. Both of these parameters are easily accessible from a PL measurement in an in-plane magnetic field (Voigt configuration), where the brightening of the $\mathrm{X_d}$ is related to a field-induced mixing of bright and dark states \cite{dark_cdse_tsmol_prl, bayer_tsmol_prl,leger_tsmol_prl}. These experiments were performed in a typical geometry, i.e., collecting the light parallel to the growth axis.

Figure \ref{fig2}(a) presents spectra of a single QD for different values of the in-plane magnetic field. Emission lines were identified as originating from recombination of neutral excitons (bright and dark), positively charged exciton ($\mathrm{X^+}$) and biexciton (2X). Figure \ref{fig2}(b) shows the time-dependent PL intensity of $\mathrm{X_d}$ transition for several values of magnetic field. The data was corrected for dark counts by subtraction of a reference signal measured at emission energy corresponding to flat PL background. Decay curves from Fig. \ref{fig2}(b) clearly show a biexponential behavior, which is due to the presence of two nearly degenerate dark excitons \cite{bayer_tsmol_prl, gershoni_tsmol_prl} with different efficiencies of field-induced mixing with bright excitons. The fast decay time is related to $\mathrm{X_d}$ with larger admixture of the bright excitons, while the slow decay time originates from PL of $\mathrm{X_d}$ with lower bright exciton admixture. This explanation stays in an agreement with more abrupt reduction of the fast decay time with increasing magnetic field. In the following analysis we focus on the dark exciton with stronger field-induced mixing. Figure \ref{fig2}(c) shows the inverse lifetime of this dark exciton as a function of the magnetic field.

We use the data in Fig. \ref{fig2}(c) to extrapolate the $\mathrm{X_d}$ lifetime to $B\! = \!0$T and thus obtain the zero-field dark exciton lifetime. To do so, we use a simple model of the neutral exciton in the in-plane magnetic field given by the following Hamiltonian \cite{bayer_tsmol_prl, leger_tsmol_prl}:
\begin{eqnarray}
\mathcal{H}&=&-2\delta_0S^z_eS^z_h+\frac{\delta_1}{2}\left(S^{+}_eS^{-}_h+S^{-}_eS^{+}_h\right)\nonumber\\
&+&g_e\mu_B\vec{B}\cdot\vec{S}_e+g_h\mu_B\vec{B}\hat{r}_{2\theta}\vec{S}_h,
\end{eqnarray}
where $\vec{S}_e$ are the electron spin operators, $\vec{S}_h$ are the 1/2 spin operators in the two-dimensional subspace of heavy hole states. The first two terms represent the isotropic and anisotropic parts of the electron-hole exchange interaction. The remaining terms represent the Zeeman energies of the electron and the hole, with their in-plane g-factors equal to $g_e$ and $g_h$, respectively. The hole Zeeman term includes the $\hat{r}_{2\theta}$ tensor, which is the rotation matrix through $2\theta$, where $\theta$ is an angle between direction of the axis of exchange interaction anisotropy and the direction related to the valence band mixing \cite{koudinov_tsmol_prl}. All the parameters in the Hamiltonian were directly extracted from the polarization resolved PL measurements in different magnetic fields for each studied QD. In particular, exchange energies were identified as the splitting between the dark and bright states ($\delta_0$), and the splitting between two bright configurations ($\delta_1$), obtained at zero magnetic field \cite{bayer_tsmol_prl}. Here we neglect $\delta_2$ splitting of the dark excitons, which is $\sim \! \mu$eV \cite{gershoni_tsmol_prl}. The carrier g-factors were extracted from the energy positions of four-fold split trion emission lines in magnetic field \cite{leger_tsmol_prl, koudinov_tsmol_prl}. Futhermore, the angle $\theta$ was obtained from the polarization resolved $\mathrm{X_d}$ and trion PL measurements \cite{koudinov_tsmol_prl}. By diagonalization of the Hamiltonian for a given magnetic field $B$, one obtains four eigenstates. Two lower energy states $|\psi_{i}(B)\rangle$ ($i=1,2$) correspond to mostly dark states. Their overlap with zero-field bright states $|\pm1\rangle$ is given by $f_{i}(B)=|\langle1|\psi_{i}(B)\rangle |^2+|\langle-1|\psi_{i}(B)\rangle |^2$. This overlap is proportional to the oscillator strength of $|\psi_{i}(B)\rangle$ radiative recombination induced by an in-plane magnetic field.  If there were no other decay mechanisms, the inverse lifetime of dark excitons would be proportional to $f_{i}(B)$ implying that at $B\! = \!0$T the $\mathrm{X_d}$ lifetime would be infinitely long. Taking into account the evidenced above field-independent radiative decay channel related to the valence band mixing, the shorter lived ($i=1$) dark exciton lifetime $\tau_1(B)$ should read
\begin{equation}
\frac{1}{\tau_1(B)}=\gamma f_{1}(B)+\frac{1}{\tau_{0}},
\label{dark_oscilator_stength}
\end{equation}
where $\tau_0$ is its zero-field lifetime, and $\gamma$ is a constant related to the radiative lifetime of the bright exciton at $B=0$T. Our calculations of Eq. (\ref{dark_oscilator_stength}) perfectly reproduce the dependence of $\mathrm{X_d}$ lifetime on the magnetic field, as it is shown in Fig. \ref{fig2}(c).
\begin{figure}
\includegraphics{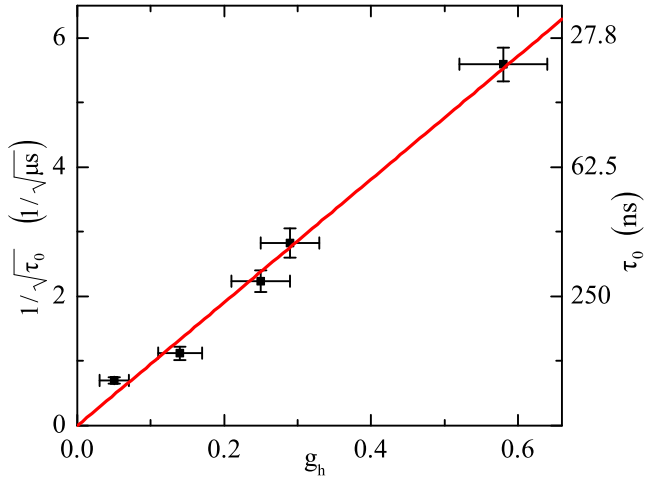}
\caption{(color online) Square root of the inverse zero-field dark exciton lifetime versus in-plane hole g-factor (each experimental point corresponds to different randomly selected QD). The solid line represents the linear fit predicted by the theory.
 \label{fig3}}
\end{figure}
In order to draw robust conclusions, similar measurements were performed on several randomly selected QDs. The obtained zero-field dark exciton lifetime $\tau_0$ varied from about $30$ns to over $2\mu$s. Figure \ref{fig3} presents the square root of $1/\tau_0$ as a function of $g_h$ for each studied QD. Within the experimental uncertainties, we observe a direct correlation between the dark exciton zero-field decay rate and the in-plane hole g-factor (i.e., the magnitude of the valence band mixing).
This clearly implies that the radiative recombination is the \emph{only important} decay channel of the dark exciton at $B=0$T in the case of CdTe/ZnTe QDs. In particular, such a result shows that the typically invoked spin-flip process turning the dark exciton into the bright one \cite{labeau_tsmol_prl, dalgarno_tsmol_prl, smith_tsmol_prl, johansen_tsmol_prl} does not play any significant role under our experimental conditions.

\begin{figure}
\includegraphics{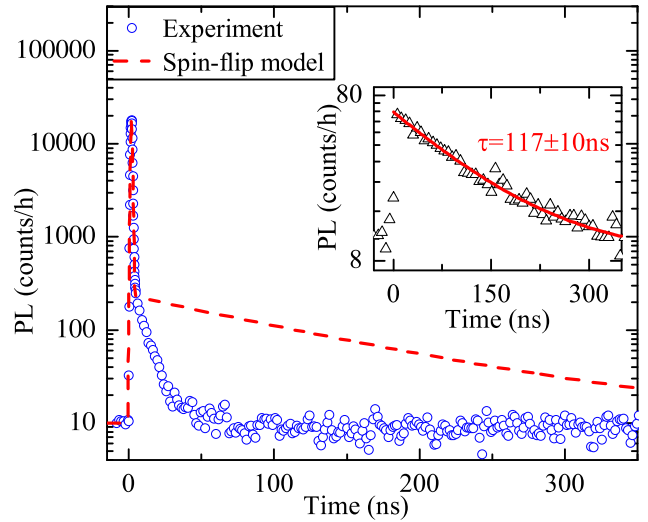}
\caption{(color online) PL decay curve of the bright exciton at $B=0$T. The dashed line represents the PL decay of the bright exciton predicted by the spin-flip model assuming equal initial populations of bright and dark states after an excitation pulse due to independent capture of electrons and holes under nonresonant excitation \cite{suffczynski_tsmol_prl, kazimierczuk_2010_tsmol_prl} (the integrals under fast and slow components appear to be different due to the semi-log scale of the plot). Zero-field lifetime of the dark exciton for this dot yielded $125\pm20$ns. Inset: the PL decay of the dark exciton for the same quantum dot at $B=0.5$T with exponential fit (solid line). For clarity, in both plots the CW background was artificially set to 10.
 \label{fig4}}
\end{figure}

Dominance of the in-plane emission of the dark exciton over the spin-flip process can be also independently ascertained by detailed analysis of the dynamics of the bright exciton recombination. Figure \ref{fig4} presents the time-resolved PL of the $\mathrm{X_b}$ measued at $B=0$T. Although $\mathrm{X_b}$ decay has the biexponential character, both lifetimes are over an order of magnitude shorter than the actual lifetime of the dark exciton in the same QD. Therefore neither fast nor slow component of the bright exciton decay can be attributed to the feeding by the spin-flip effect. We identify the faster $\mathrm{X_b}$ decay time with the radiative lifetime of the bright exciton. The slower decay, which corresponds to about 5\% of the total line intensity, is tentatively attributed to the recapturing of the carriers by the QD. Qualitative discrepancy between the experimental $\mathrm{X_b}$ decay and the one calculated in the spin-flip model (dashed line in Fig. \ref{fig4}) clearly shows that the spin-flip process is ineffective in the case of CdTe/ZnTe QDs.

All presented results are a clear evidence for the radiative recombination channel of the dark exciton. This radiative recombination can dominate over other decay mechanisms of the dark exciton. We believe that our findings will stimulate the studies on the role of the dark exciton in the QDs. In particular, the coupling between the dark exciton and the in-plane radiation demonstrated in our work may be used for a direct optical control of the dark exciton qubit with lifetime much longer than a qubit based on the bright exciton in a QD.

This work was partially supported by the Polish Ministry of Science and Higher Education in years 2012-2016 as research grants ``Iuventus" and a ``Diamentowy Grant", and by the National Science Centre under Grants No. DEC-2011/02/A/ST3/00131 and No. DEC-2011/01/N/ST3/04536. Experiments were carried out with the use of CePT, CeZaMat and NLTK infrastructures financed by the European Union - the European Regional Development Fund within the Operational Programme ``Innovative economy'' for 2007-2013. \L{}C acknowledges support from the Homing programme of the Foundation for Polish Science supported by the EEA Financial Mechanism.


\begin{thebibliography}{99}

\bibitem{loss_tsmol_prl}
D. Loss and D. P. DiVincenzo, Phys. Rev. A \textbf{57}, 120 (1998).

\bibitem{liu_tsmol_prl}
R.-B. Liu, W. Yao, and L. J. Sham, Adv. Phys. \textbf{59}, 703 (2010).

\bibitem{single_photon_tsmol_prl}
P. Michler, A. Kiraz, C. Becher, W. V. Schoenfeld, P. M. Petroff, Lidong Zhang, E. Hu, and A. Imamo\u{g}lu, Science \textbf{290}, 2282 (2000).

\bibitem{entangled_photon1_tsmol_prl}
N. Akopian, N. H. Lindner, E. Poem, Y. Berlatzky, J. Avron, D. Gershoni, B. D. Gerardot, and P. M. Petroff, Phys. Rev. Lett. \textbf{96}, 130501 (2006).

\bibitem{entangled_photon2_tsmol_prl}
R. M. Stevenson, R. J. Young, P. Atkinson, K. Cooper, D. A. Ritchie, and A. J. Shields, Nature (London) \textbf{439}, 179 (2006).

\bibitem{labeau_tsmol_prl}
O. Labeau, P. Tamarat, and B. Lounis, Phys. Rev. Lett. \textbf{90}, 257404 (2003).

\bibitem{suffczynski_tsmol_prl}
J. Suffczy\'nski, T. Kazimierczuk, M. Goryca, B. Piechal, A. Trajnerowicz, K. Kowalik, P. Kossacki, A. Golnik, K. P. Korona, M. Nawrocki, J. A. Gaj, and G. Karczewski, Phys. Rev. B \textbf{74}, 085319 (2006).

\bibitem{gershoni_tsmol_prl}
E. Poem, Y. Kodriano, C. Tradonsky, N. H. Lindner, B. D. Gerardot, P. M. Petroff, and D. Gershoni, Nature Physics \textbf{6}, 993 (2010).

\bibitem{dark_cdse_tsmol_prl}
M. Nirmal, D. J. Norris, M. Kuno, M. G. Bawendi, Al. L. Efros, and M. Rosen, Phys. Rev. Lett. \textbf{75}, 3728 (1995).

\bibitem{bayer_tsmol_prl}
M. Bayer, G. Ortner, O. Stern, A. Kuther, A. A. Gorbunov, A. Forchel, P. Hawrylak, S. Fafard, K. Hinzer, T. L. Reinecke, S. N. Walck, J. P. Reithmaier, F. Klopf, and F. Sch\"afer, Phys. Rev. B \textbf{65}, 195315 (2002).

\bibitem{goryca_tsmol_prl}
M. Goryca, P. Plochocka, T. Kazimierczuk, P. Wojnar, G. Karczewski, J. A. Gaj, M. Potemski, and P. Kossacki, Phys. Rev. B \textbf{82}, 165323 (2010).

\bibitem{dalgarno_tsmol_prl}
P. A. Dalgarno, J. M. Smith, B. D. Gerardot, A. O. Govorov, K. Karrai, P. M. Petroff, and R. J. Warburton, Phys. Stat. Sol. (a) \textbf{202}, 2591 (2005).

\bibitem{smith_tsmol_prl}
J. M. Smith, P. A. Dalgarno, R. J. Warburton, A. O. Govorov, K. Karrai, B. D. Gerardot, and P. M. Petroff, Phys. Rev. Lett. \textbf{94}, 197402 (2005).

\bibitem{roszak_tsmol_prl}
K. Roszak, V. M. Axt, T. Kuhn, and P. Machnikowski, Phys. Rev. B \textbf{76}, 195324 (2007).

\bibitem{johansen_tsmol_prl}
J. Johansen, B. Julsgaard, S. Stobbe, J. M. Hvam, and P. Lodahl, Phys. Rev. B \textbf{81}, 081304(R) (2010).

\bibitem{tinjod_tsmol_prl}
F. Tinjod, B. Gilles, S. Moehl, K. Kheng, and H. Mariette, Appl. Phys. Lett. \textbf{82}, 4340 (2003).

\bibitem{kazimierczuk_2010_tsmol_prl}
T. Kazimierczuk, M. Goryca, M. Koperski, A. Golnik, J. A. Gaj, M. Nawrocki, P. Wojnar, and P. Kossacki, Phys. Rev. B \textbf{81}, 155313 (2010).

\bibitem{kazimierczuk_2011_tsmol_prl}
T. Kazimierczuk, T. Smole\'nski, M. Goryca, \L{}.~K\l{}opotowski, P. Wojnar, K. Fronc, A. Golnik, M. Nawrocki, J. A. Gaj, and P. Kossacki, Phys. Rev. B \textbf{84}, 165319 (2011).

\bibitem{leger_tsmol_prl}
Y. L\'eger, L. Besombes, L. Maingault, and H. Mariette, Phys. Rev. B \textbf{76}, 045331 (2007).

\bibitem{karlson_tsmol_prl}
K. F. Karlsson, M. A. Dupertuis, D. Y. Oberli, E. Pelucchi, A. Rudra, P. O. Holtz, and E. Kapon, Phys. Rev. B \textbf{81}, 161307(R) (2010).

\bibitem{dupertuis}
M. A. Dupertuis, K. F. Karlsson, D. Y. Oberli, E. Pelucchi, A. Rudra, P. O. Holtz, and E. Kapon, Phys. Rev. Lett. \textbf{107}, 127403 (2011).

\bibitem{krizankovski_tsmol_prl}
D. N. Krizhanovskii, A. Ebbens, A. I. Tartakovskii, F. Pulizzi, T. Wright, M. S. Skolnick, and M. Hopkinson, Phys. Rev. B \textbf{72}, 161312(R) (2005).

\bibitem{atature_tsmol_prl}
M. Atat\"ure, J. Dreiser, A. Badolato, A. H\"ogele, K. Karrai, and A. Imamo\u{g}lu, Science \textbf{312}, 551 (2006).

\bibitem{ohno_tsmol_prl}
S. Ohno, S. Adachi, R. Kaji, S. Muto, and H. Sasakura, Appl. Phys. Lett. \textbf{98}, 161912 (2011).

\bibitem{koudinov_tsmol_prl}
A. V. Koudinov, I. A. Akimov, Yu. G. Kusrayev, and F. Henneberger, Phys. Rev. B \textbf{70}, 241305(R) (2004).

\bibitem{cao_tsmol_prl}
C. L. Cao, L. Besombes, and J. Fern\'andez-Rossier, Phys. Rev. B \textbf{84}, 205305 (2011).

\end{thebibliography}
\end{document}